\documentstyle[multicol,aps]{revtex}
\input{psfig}
\begin{document}
\pagestyle{myheadings}
\draft
\title{Scaling of Low-Order Structure Functions 
in Homogeneous Turbulence}
\author{Nianzheng Cao$^{1,2}$, Shiyi Chen$^{1,2}$ and
 Katepalli R. Sreenivasan$^{3}$}
\address{${}^{1}$IBM Research Division, T. J. Watson Research Center,
P.O. Box 218, Yorktown Heights, NY 10598\\
${}^{2}$Theoretical Division and Center for Nonlinear Studies,
Los Alamos National Laboratory, Los Alamos, NM 87545\\
${}^{3}$Mason Laboratory, Yale University, New Haven, CT 06520-8286}

\maketitle

\begin{abstract}

\noindent High-resolution direct numerical simulation data for  
three-dimensional Navier-Stokes turbulence in a periodic box
are used to study the scaling behavior of low-order 
velocity structure functions with positive and negative powers. 
Similar to high-order statistics, the low-order relative scaling  
exponents exhibit unambiguous departures from the Kolmogorov 1941 
theory and agree well with existing multiscaling models. No transition 
from normal scaling to anomalous scaling is observed.

\end{abstract}
\pacs{47.27.-i, 47.27.Gs}
\begin{multicols}{2}
\narrowtext

The intermittency effect of small scale dynamics on fluid turbulence 
has been an important issue in turbulence statistics for many  
decades. A lot of effort has been invested towards the understanding 
of the  departure from Gaussian statistics and the related anomalous 
scaling behavior of velocity structure functions. 
At very high Reynolds numbers, 
Kolmogorov's 1941 similarity theory (K41) \cite{k41}
predicts a normal (or regular) scaling for moments of 
the velocity increment in the inertial range:
$\langle |\Delta u_r|^q \rangle \sim r^{q/3}$ for $L >> r >> \eta$,  
where the longitudinal velocity increment is defined as 
$\Delta u_r = u(x+r) - u(x)$,
$\langle \cdot \rangle$ denotes an ensemble average and $L$
and $\eta$ are, respectively, the integral scale and the Kolmogorov 
dissipation scale. For convenience, we have considered here the  
moments of the absolute values of velocity differences --- called the  
``generalized structure functions" to distinguish them from the  
classical ones. The normal scaling has been challenged often and the  
departure from K41 has been reported from both experiments and direct
numerical simulations \cite{k62,sreeni1,kuo}. 
A recent paper \cite{cao} used data from a high-resolution simulation  
and demonstrated the existence of intermittency correction to K41  
{\it without} invoking the Taylor's hypothesis used in real-life  
experiments.

It has been believed traditionally that the deviations from 
K41 tend to be more pronounced when the moment order $q$ becomes  
large. This is because the non-Gaussianity is more conspicuous at 
the tails of the probability density function (PDF) which
make increasingly significant contributions to higher order  
statistics. Thus, experiments and 
simulations have focused on structure functions with
$q \geq 2$, say. With the exception of References \cite{sreeni2} and
\cite{kadanoff}, little attention has been paid to the scaling of low 
order structure functions ($q < 2$, say). The preliminary experimental 
results of a water flow in a pipe in Ref. \cite{sreeni2} showed that exponents 
for moment orders as low as 0.25 departed from K41. However, a detailed 
study connecting large amplitude events with structure functions and 
scaling exponents has not yet been made.

For one-dimensional Burgers turbulence in the small viscosity limit,  
a bifractal structure of velocity due to the formation of shock
structures leads to a normal scaling for $q<1$ and a constant scaling 
exponent for $q \geq 1$ \cite{cheklov}. This bifractal scaling has  
also been reported for turbulent convection with a constant 
temperature gradient \cite{vadim}. One cannot in principle rule out a  
somewhat similar situation in three-dimensional turbulence at very  
high Reynolds numbers. It is therefore both interesting and useful to  
examine the nature of low-order scaling exponents.

In this Letter, we present an analysis using data from direct numerical 
simulations for a fully developed isotropic turbulence to study low order moments 
of velocity structure functions and the relation
between structure function exponents and the PDF. 
The particular questions we address here are: (1) Are there
intermittency corrections to the scaling of low order structure  
functions? (2) Is there a transition from regular scaling at low 
order to anomalous scaling at high order? (3) What aspects of 
turbulence make major contribution to low order structure 
function statistics?

Direct numerical simulation of the Navier-Stokes equations  
\cite{chen} was carried out with $512^3$ mesh points in a 
cyclic cubic box for homogeneous isotropic turbulence using 
the CM-5 machine at Los Alamos and the
SP machines at IBM. The simulation domain was [0, $2\pi$] in each  
direction. A nominal steady state was maintained by 
a forcing confined to wavenumbers $k < 3$.  
The Taylor microscale $\lambda=(15\nu v_ 0^2/\epsilon)^{1/2}$ and  
microscale Reynolds number ${\cal R}_\lambda \equiv v_0\lambda/\nu$ were  
controlled by varying  the viscosity of the system. Here $v_0$ is the 
root-mean-square value of a single vector component of velocity. 
The analysis was carried out for forced statistically 
steady states at ${\cal R}_\lambda = 218$. A spatial averaging over  
the whole
physical space was used to replace the ensemble average. The  
separation $r$ was 
taken along the $x$ direction and the PDFs and statistics shown below 
were averaged over 1.5 large-eddy turnover times.

In the inset of Fig. 1, we present the generalized structure  
functions
\begin{equation}
S_q(r)=\langle|\Delta u_r|^q\rangle \sim r^{\zeta_q} , 
\end{equation}
with $q=0.4$ and $0.8$. Here $\zeta_q$ is the scaling exponent.
A power law range can be identified for   
$0.2\leq r \leq 0.8$. In Ref. \cite{sreeni2}, we have shown that the  
third-order structure function in this region has the expected  
inertial range behavior \cite{k41b}. The structure functions are  
compared with a phenomenological multiscaling model proposed by She  
and Leveque \cite{she}
(solid lines). To better calculate the scaling exponents, in Fig. 1 we have  
utilized the so-called
extended scaling similarity (ESS) hypothesis advocated by
Benzi {\em et al.} \cite{benzi}, i.e., plotting $S_q(r)$ against  
$S_3(r)$, and identified the relative scaling exponent, $\zeta_q$. 
From our previous experiences \cite{cao} and the current 
study, we believe that the ESS works well for velocity structure
functions in homogeneous turbulence---meaning that the relative  
scaling shows a wider region than the direct scaling region shown in  
the inset.  We have studied many instances of direct
and relative scalings for different power indexes \cite{fn},
ranging from $q=-0.8$ to $12$, and found the scaling relations to be  
very similar to those presented in Fig. 1.

\bigskip
\psfig{file=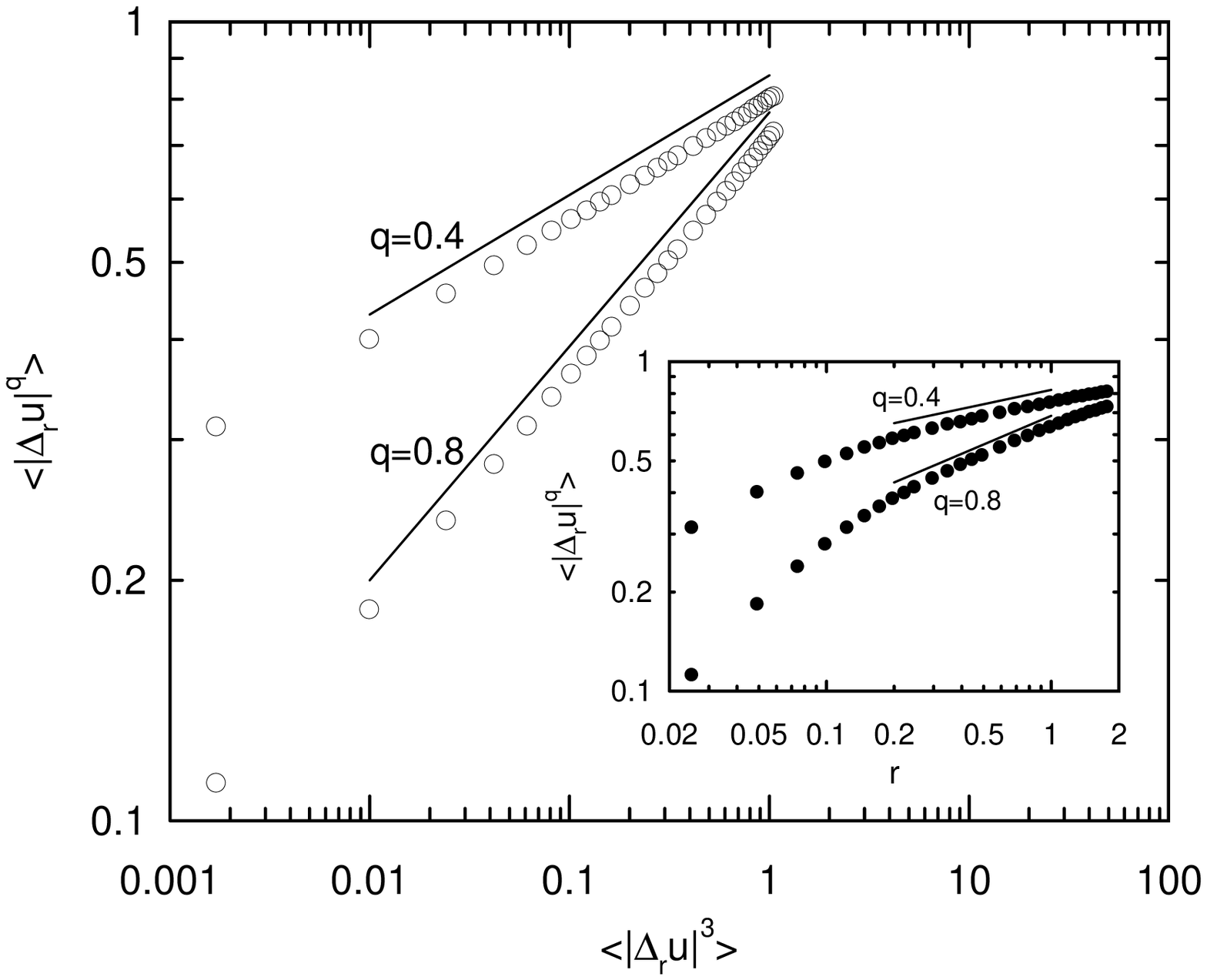,width=234pt}
\noindent
 {\small FIG.~1. Velocity structure function
$S_q(r)$ for $q=0.4$ and 0.8 versus $S_3(r)$. The inset shows
the velocity structure functions versus $r$. The symbols are
direct numerical simulation results, and solid lines indicate
slopes from the SL model.}
\bigskip

In Fig. 2 we have compared the ESS scaling exponents for $-0.8<q<2$
with predictions from K41, a log-normal model by
Kolmogorov (K62) \cite{k62} which gives $\zeta_q = q/3+q(3-q)\mu/18$ 
where $\mu=2/9$ is the dissipation
scaling exponent \cite{sreeni3}, and a log-Poisson model by She and  
Leveque (SL) \cite{she} which gives $\zeta_q = q/9 + 2/3[1-(2/3)^{q/3}]$. 
To be more quantitative, we list here five scaling exponents from 
direct numerical simulations: 
$\zeta_{-0.8}=-0.32\pm0.01$, $\zeta_{0.2}=0.074\pm0.002$,
$\zeta_{0.4}=0.150\pm0.002$, $\zeta_{0.6}=0.223\pm0.002$,
$\zeta_{0.8}=0.296\pm0.002$.
For very small $q$, $\zeta_q$  
should be a linear function of $q$ following a Taylor expansion.
The difference of the slope of scaling
exponents when $q\rightarrow 0$ between K41 and SL model (or K62)
is about $12\%$ \cite{sreeni1}. It is evident that the scaling  
exponents deviate from K41 even for very small $q$, in favor of the
intermittency models \cite{k62,she,chen1} for both positive 
and negative powers.  Since the departure from K41 starts from very 
small $q$ and the relative scaling exponents agree well with some intermittency 
models for $q$ from $-0.8$ up to 8 \cite{cao}, there is no evidence to  
support a transition of scaling exponents and the bifractal structure,  
in contrast to the case of Burgers 
equation \cite{cheklov} or pressure-less turbulence \cite{polyakov}.

\bigskip
\psfig{file=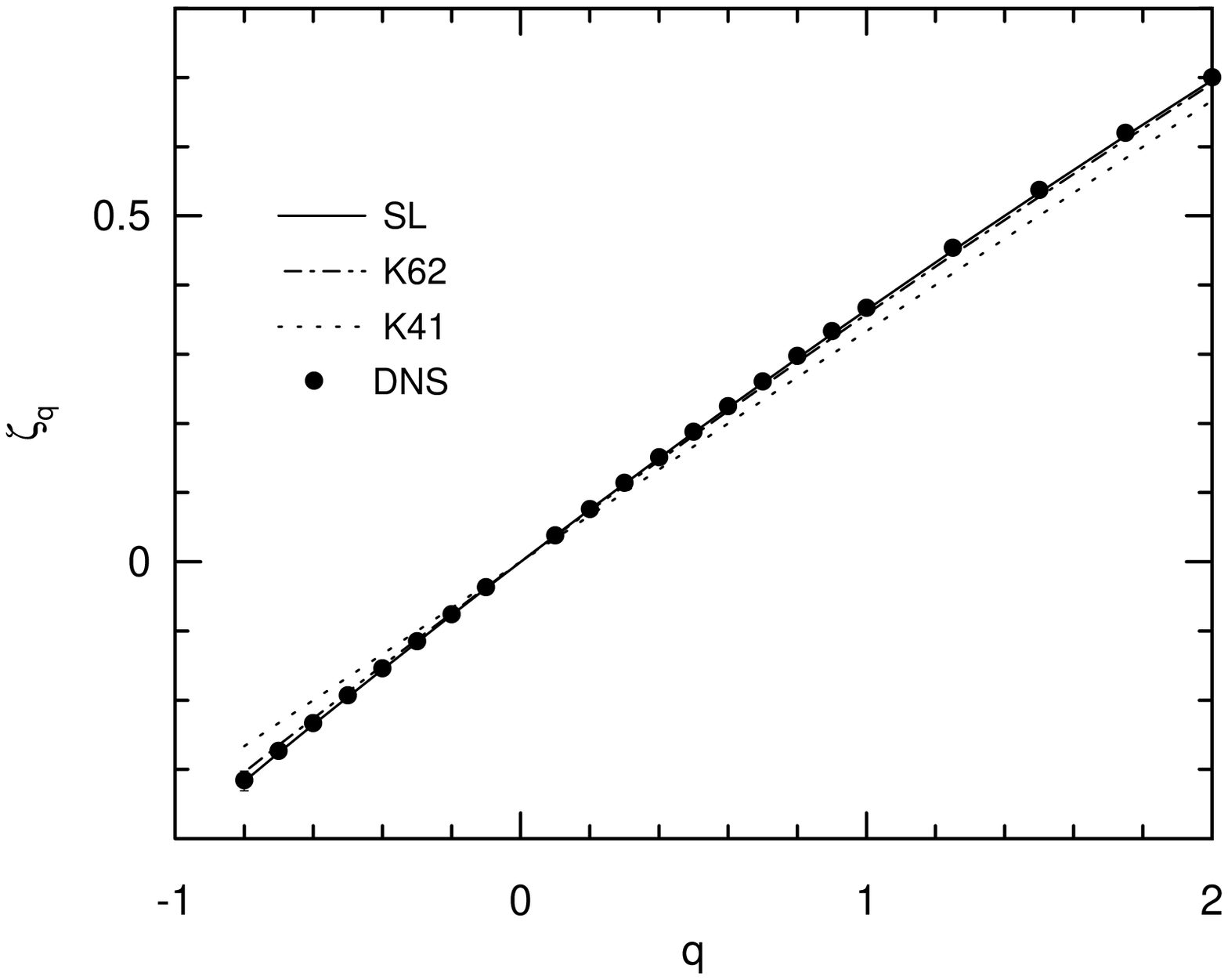,width=234pt}
\noindent
{\small FIG.~2. Scaling exponents of structure functions as a
function of the order index $q$ compared with various models.}
\bigskip

As pointed out in \cite{sreeni1}, it is generally believed that
the PDF of the velocity increment has a Gaussian core for small  
amplitude events and a stretched exponential shape for the tail part.
The intermittency effect presented in Fig. 2 for
low order moments may imply two possibilities: (1) the low order  
moments have substantial contributions from the high amplitude (tail)  
events; (2) the core part of the PDFs is also sufficiently non-Gaussian. 
To  shed light on these possibilities, we have enlarged in Fig. 3 the  
core part PDF of $\Delta u_r$ with the inset showing the whole PDF. 
While the  deviation from the Gaussian is evident at the tails, 
even the core  part of the PDF displays departures from Gaussianity. 
This departure may have significant contributions to low order 
statistics. 

\bigskip
\psfig{file=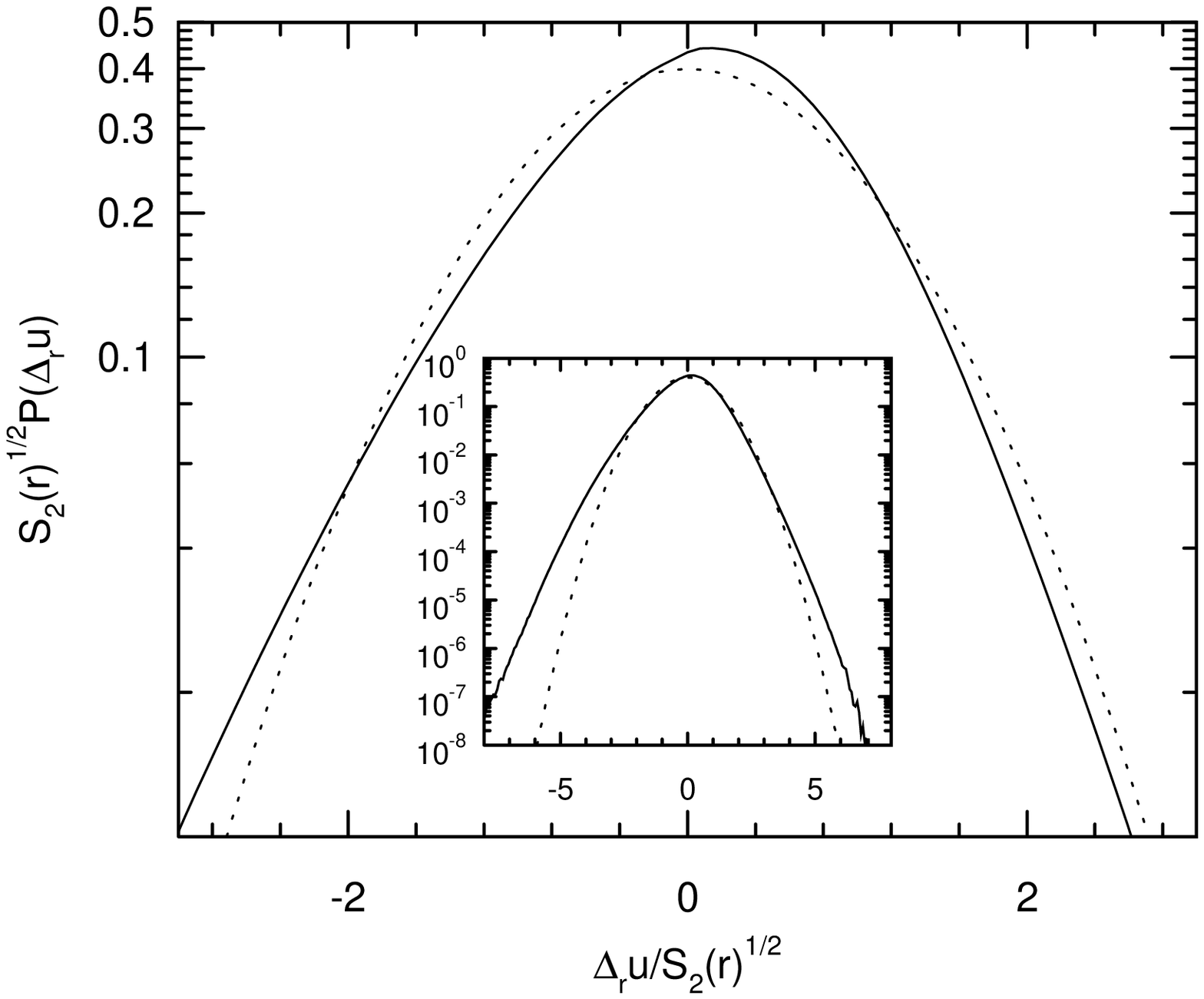,width=234pt}
\noindent
{\small FIG.~3. The core part of the PDF for velocity increment
(solid line) compared with Gaussian distribution (dotted line).
The inset shows the whole distribution for the PDFs. All curves are
normalized by the root-mean-square.}
\bigskip

To probe this further, we define a cumulative structure function,
\begin{equation}
S_q(r,\phi)  = \int_{-\phi}^{\phi}|\phi'|^qP(\phi')d\phi',
\end{equation}
and the cumulation ratio
\begin{equation}
\gamma_q(\phi)=\frac{S_q(r,\phi)}{S_q(r)}
\end{equation}
where $\phi'=\Delta u_r/S_{rms}(r)$ and  $S_{rms}(r) = S_2(r)^{1/2}$ 
is the root-mean-square for a given separation $r$; $P(\phi')$ is the  
probability density function of $\phi'$; $\phi$ is the bound of truncated
$\phi'$ domain. The ratio
$\gamma_q(\phi)$ {\em quantitatively} defines a cumulative  
contribution from events with amplitude up to $\phi$ for the 
$q^{th}$-order structure function.  In Fig. 4, we show 
$\gamma_q(\phi)$ as a function of $\phi$
for a separation distance in the inertial range ($r=0.29$). For fixed  
$\gamma_q(\phi)$, we see that the larger the
$q$  the larger the $\phi$, consistent with the fact that the peak in  
$\phi^q P(\phi)$ shifts to larger values of $\phi$ for increasingly  
high-order moments.
We provide in the inset the complementary data on
$\phi$ (the inverse function of $\gamma$ in Eq. (3))
as a function of $q$ when $\gamma_q(\phi)=0.9$.

\bigskip
\psfig{file=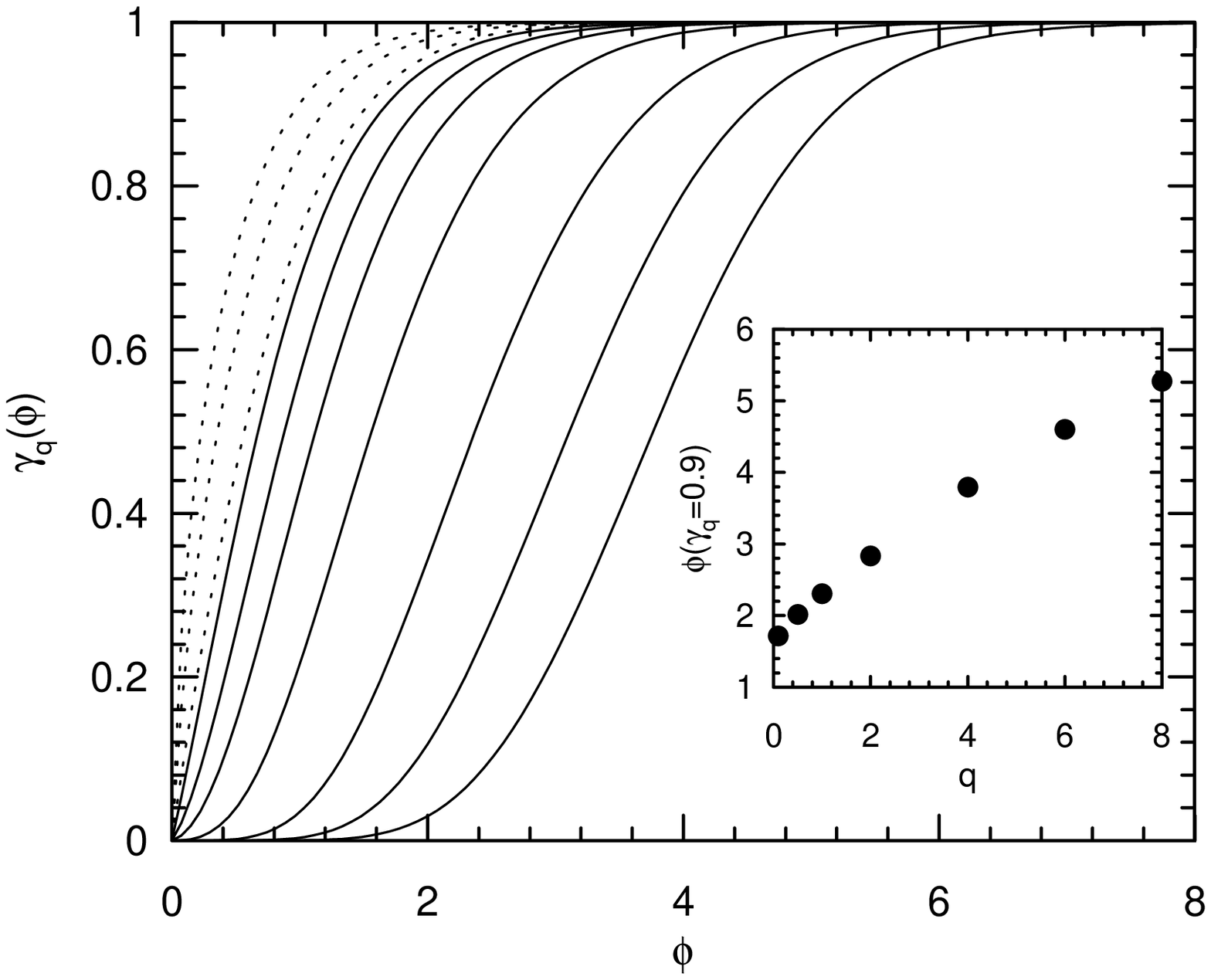,width=234pt}
\noindent
{\small FIG.~4.  The cumulation ratio $\gamma_q(\phi)$ as functions of
$\phi$ for different $q$ (from left to right): $q=-0.8$, $-0.5$, $-0.1$,
$0.1$, $0.5$, $1$, 2, 4, 6, 8. The separation distance $r=0.29$ lies
within the inertial range. The inset shows $\phi$ values
corresponding to the intersections of the curves by the horizontal
line given by $\gamma_q = 0.9$.}
\bigskip

The aspect just discussed has been made more specific in Fig. 5 for  
the flatness and skewness of the velocity increment ($r=0.29$).
The data are compared with those of a Gaussian field. The departure  
from Gaussian statistics for the flatness starts from $\phi \simeq  
1$, which might therefore usefully 
define the boundary between the core and the tail of the 
PDF. Both the skewness and the flatness converge to constants
(3.6 and -0.32 respectively) when $\phi\simeq 4$, consistent
with the inset in Fig. 4. It is interesting to note that the skewness
has a crossover from positive to negative at $\phi=2$, indicating  
that the negative skewness basically comes from events of intermediate  
amplitudes between about 2 and 4 standard deviations.

\bigskip
\psfig{file=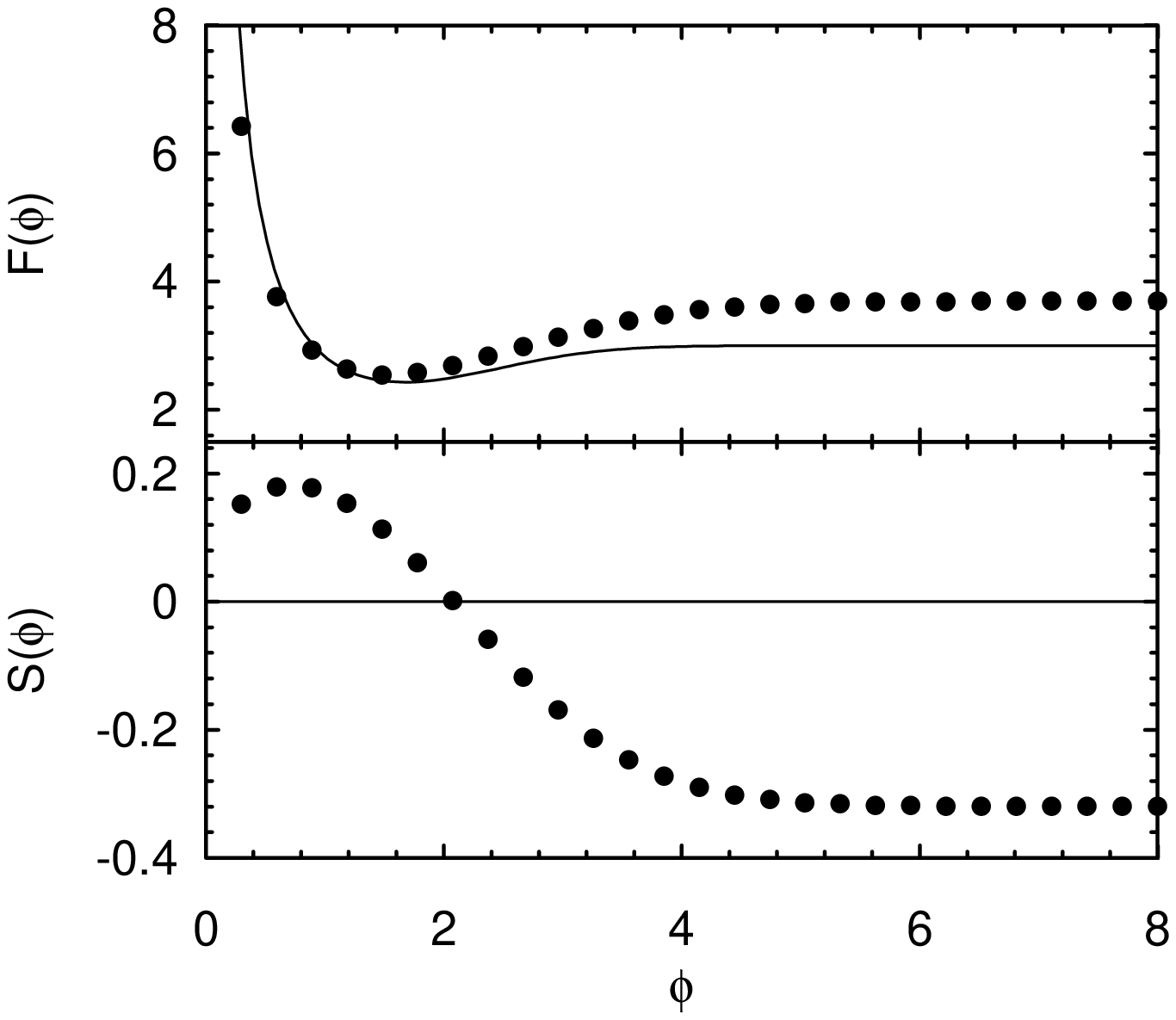,width=234pt}
\noindent
{\small FIG.~5. The skewness and flatness factor as function of
$\phi$ for $r=0.29$ calculated from the data (bullets), compared with
Gaussian values (solid lines).}
\bigskip

In order to see the effects of the PDF on scaling exponents, in
particular on the variation of scaling exponents with the variation
of the upper bound, $\phi$, it is useful to rewrite Eq. (1) as
$\zeta_q(r)= d \ln{S_q(r)}/d\ln{r}$. We first note that the
scaling exponents represent a {\it relative variation} of the
structure function as a function of the separation distance. As in
Eq. (2), we can now define the $q^{th}$-order cumulative scaling
exponent as 
\[ 
\zeta_q(r,\phi)=\frac{d\ln{S_q(r,\phi)}}{d\ln{r}}. 
\]
In Fig. 6, we show $\zeta_q(\phi)$, the average of $\zeta_q(r,\phi)$
over inertial range, as a function of $\phi$ 
for $q=0.4,~2$ and $8$. Two features in this plot
are worth noting: First, for $\phi \simeq 1$, the scaling exponents
for all three cases are close to K41. Second, all three cases 
converge to the SL multiscaling model when $\phi\simeq 6$.
This result implies that---as far as the scaling exponents are  
concerned---there is no fundamental difference 
in the relation between structure function and the PDF of 
the velocity increment for high order and low order statistics. 
Note that the transition in Fig. 6 from K41 to the SL 
(or other) multiscale results is smooth, which
seems to support the multifractal picture in which the
correction to anomalous scaling comes from all amplitudes with  
different fractal dimensions.

\bigskip
\psfig{file=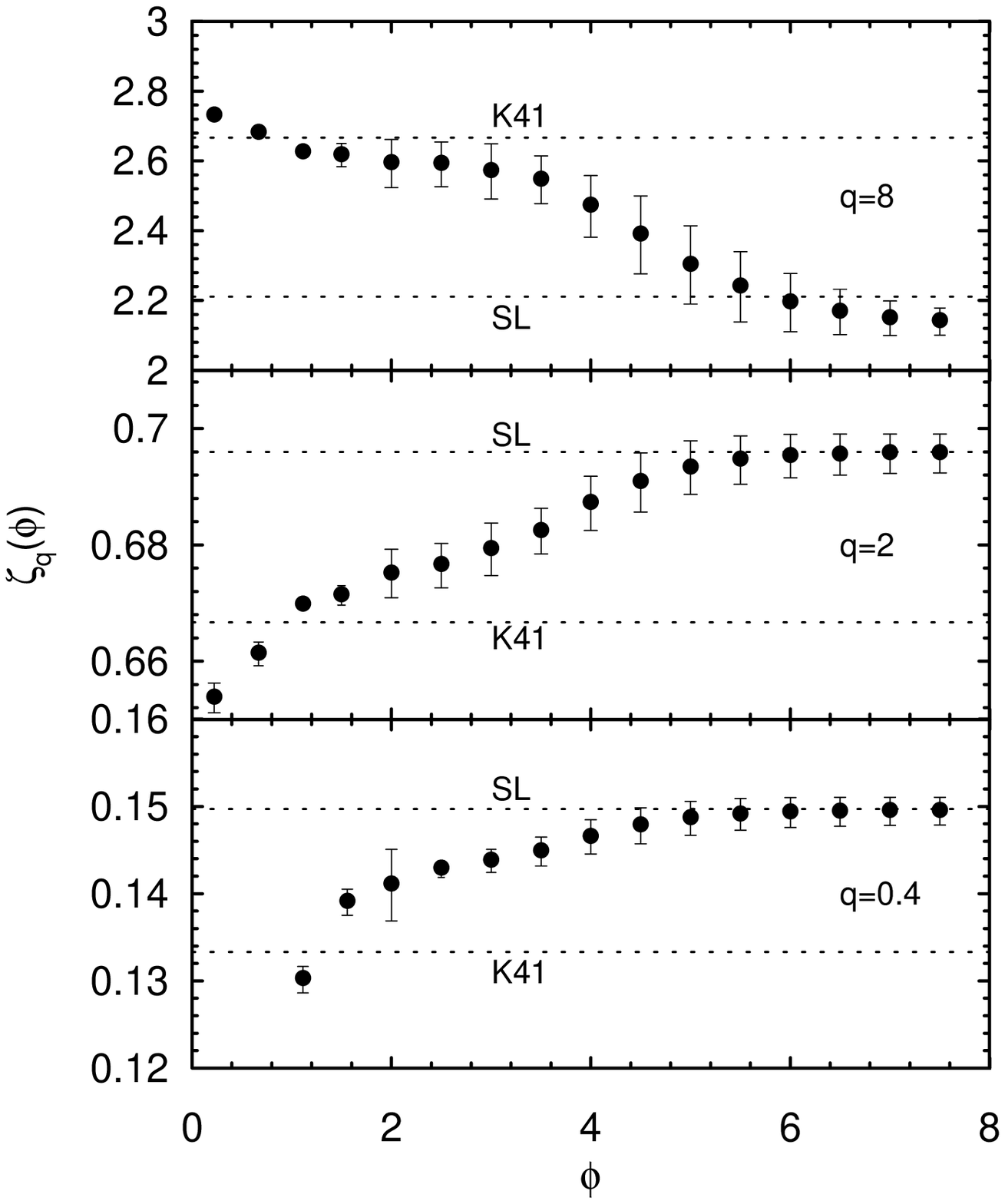,width=234pt}
\noindent
{\small FIG.~6. Scaling exponents $\zeta_q(\phi)$ versus
$\phi$ for $q=0.4, 2$ and $8$ for truncated structure functions.
The values for complete structure functions from K41 and
SL are also plotted for comparison.}
\bigskip

It is not often appreciated that the non-Gaussianity of PDF of  
velocity increments is related to the phase coherence among velocity  
mode in Fourier space (``structure" in real space): when a  
pseudo-velocity field is generated by randomizing the phase for each  
mode (but not altering its amplitude) the resulting scaling
exponents agree with K41. Although this result is not surprising, it  
confirms
the direct link of intermittency and anomalous scaling to real space  
structures \cite{she1}. 

To summarize, we have reported a study of low order scaling 
for the direct numerical simulation data on isotropic turbulence. 
The scaling exponents for low order structure functions clearly 
deviate from K41 but agree well with existing multiscaling models. 
No transition from regular scaling to anomalous
scaling can be identified. The dependence of structure functions
and scaling exponents on amplitudes of the velocity increment are 
investigated. While the low and high order structure functions have 
primary contributions from small and large amplitude events,  
respectively, the scaling exponents for structure functions, 
regardless of the power index $q$, seems to have some contribution 
from all events. \\

We thank R. H. Kraichnan, D. Lohse, Z.-S. She and S.I. Vainshtein for useful 
discussions.  Part of this work was supported by the U.S. Department 
of Energy at Los Alamos. Numerical simulations 
were carried out at the Advanced Computing Laboratory at Los Alamos
using the Connection Machine-5
and the Center for Scalable Computing Solution at IBM using the SP 
machines. KRS was supported partially by the Sloan Foundation.

%
%
%
%
%
%
%
 
 \end{multicols}


\begin{references}
\bibitem {fn}
It should be pointed out that moments of negative order can only be
defined for $q > - 1$: writing structure functions as  
$S_q(r)=\int_{-\infty}^{\infty}|\Delta u_r|^q P(\Delta u_r) d\Delta  
u_r$, 

where $P(\Delta u_r)$ is the probability density function of
velocity increment $\Delta u_r$, it is easily seen---since $P(0)$ is  
a non-zero constant---that the above integral converges only when $q  
> -1$.
\end{references}
\end{document}